\RequirePackage[2020-02-02]{latexrelease}

\documentclass[twocolumn,preprintnumbers]{revtex4}%
\usepackage{amssymb}
\usepackage{amsmath}
\usepackage{graphicx}
\usepackage{epstopdf}
\usepackage{dcolumn}
\usepackage{bm}
\usepackage{xcolor}
\usepackage{ifthen}
\usepackage[normalem]{ulem}
\usepackage{amsfonts}%
\RequirePackage[2020-02-02]{latexrelease}
\graphicspath{{c:/Users/Eyal/MyDocuments/latex/Texfiles/Diamond/dipolarBA/EntangSupp/SDIP/}
	{c:/Users/Eyal/MyDocuments/latex/Texfiles/Diamond/dipolarBA/EntangSupp/PP/SCC/}
	{c:/Users/Eyal/MyDocuments/latex/Texfiles/Diamond/dipolarBA/EntangSupp/Q/}}
\UseRawInputEncoding
\providecommand{\U}[1]{\protect\rule{.1in}{.1in}}
\providecommand{\U}[1]{\protect\rule{.1in}{.1in}}
\def\showal{1}
\newcommand{\al}[1]{\ifthenelse{\showal=1}{\textcolor{orange}{[[#1]]}}{}}

\newcommand{\eb}[1]{\ifthenelse{\showal=1}{\textcolor{cyan}{[[#1]]}}{}}
\begin{document}
\title{Disentanglement in the macroscopic limit}
\author{Eyal Buks}
\email{eyal@ee.technion.ac.il}
\affiliation{Andrew and Erna Viterbi Department of Electrical Engineering, Technion, Haifa
32000, Israel}
\date{\today }

\begin{abstract}
The recently proposed spontaneous disentanglement hypothesis is formulated
using a modified Schr\"{o}dinger equation having an added nonlinear term. The
hypothesis is motivated by some outstanding issues in the foundations of
quantum mechanics, including the problem of quantum measurement. Spontaneous
disentanglement is explored in the current study for the macroscopic limit.
This is done using some many--body models having known exact solutions. For
the under--study models, it is found that non--local entanglement becomes
unstable in the macroscopic limit. On the other hand, stability in the
macroscopic limit of local entanglement is not excluded. These findings
demonstrate that the spontaneous disentanglement hypothesis can bridge between
the quantumness of the microscopic realm, and the classicalness of the
macroscopic one.

\end{abstract}
\maketitle

%Force line breaks with \\

%Lines break automatically or can be forced with \\

%It is always \today, today,
%but any date may be explicitly specified

%PACS, the Physics and Astronomy
%Classification Scheme.
%\keywords{Suggested keywords}%Use showkeys class option if keyword
%display desired

\textbf{Introduction} -- Unitary time evolution of a quantum state vector
$\left\vert \psi\right\rangle $ is governed by the Schr\"{o}dinger equation.
It is commonly assumed that the unitary time evolution is supplemented by two
auxiliary processes. One is thermalization, and the other is a collapse of the
state vector \cite{Schrodinger_807,Penrose_4864}, which occurs when a
measurement is performed. The collapse gives rise to disentanglement between a
measured quantum system and its measuring apparatus
\cite{von_Neumann_Mathematical_Foundations}. The Schr\"{o}dinger equation is
linear in $\left\vert \psi\right\rangle $, whereas both auxiliary processes
violate the principle of superposition, and consequently nonlinear equations
of motion are needed to describe them \footnote{For
the process of thermalization, nonlinearity is needed because the entropy is a
nonlinear function of the density operator. Nonlinearity is needed for
disentanglement because the subset of disentangled states is generally not a
subspace of the system's Hilbert space (see also Ref. \cite{Genoni_032303}).}. Moreover, in contrast to the time--reversibility
of the Schr\"{o}dinger equation, both auxiliary processes are
time--irreversible. The inability to describe both thermalization
\cite{Grabert_161,Ottinger_026121} and disentanglement
\cite{Bassi_471,Pearle_857,Kowalski_1,Fernengel_385701,Carlesso_243,Donadi_74}
processes using the Schr\"{o}dinger equation has motivated the study of a
variety of nonlinear extensions to quantum mechanics
\cite{Weinberg_61,Doebner_3764,Gisin_5677,Gisin_2259,Kaplan_055002,Munoz_110503,Geller_2200156,Ghirardi_470,Oppenheim_041040,Bennett_170502,Czachor_4122,Schrinski_133604}%
.

A nonlinear extension to quantum mechanics, which is based on the
hypothesis that disentanglement spontaneously occurs in quantum systems, has
been recently proposed \cite{Buks_2400036}. Equations of motion, which are
derived from this hypothesis, incorporate unitary time evolution with both
auxiliary processes of thermalization and disentanglement. It was shown that
this proposed nonlinear extension can account for some phenomena that are
experimentally observed in quantum systems, including phase transitions,
dynamical multi--stabilities and superconductivity
\cite{Buks_012439,Buks_2400587,Buks_630}. These phenomena are commonly
accounted for by employing the mean field approximation. However, the validity
of this approximation is arguably questionable, in particular when it turns
linear dynamics into nonlinear ones \cite{Barthel_052224}.

Quantum entanglement has been conclusively observed in all known microscopic
physical systems (e.g. elementary particles, atoms and molecules), and in many
mesoscopic systems. On the other hand, the applicability of standard quantum
mechanics for macroscopic systems is arguably confutable
\cite{Leggett_857,Bell_33,Adler_135}. In the current study, the proposed
nonlinear equations of motion are analyzed for some many--body quantum models
having known analytical solutions. In particular, the dependency on system
size is studied, in order to explore whether the proposed nonlinear extension
can describe the transition between quantumness in the microscopic limit, and
classicalness in the macroscopic one. Our results demonstrate that under
appropriate conditions, disentanglement can have a significant impact in the
macroscopic limit, even when its effect in the microscopic limit is made
arbitrarily small.

\textbf{Spontaneous disentanglement hypothesis} -- A nonlinear
extension to quantum dynamics can be formulated using nonlinear Kraus
operators
\cite{Elben_200501,Sergi_1350163,Brody_230405,Kaplan_055002,Geller_2200156}.
Consider the case where, to first order in the time interval $\tau$, the
density operator $\rho$\ evolves according to
\begin{equation}
\rho\left(  t+\tau\right)  =K_{0}^{{}}\rho\left(  t\right)  K_{0}^{\dag}%
+K_{1}^{{}}\rho\left(  t\right)  K_{1}^{\dag}+O\left(  \tau^{2}\right)  \;,
\label{rho(t+tau)}%
\end{equation}
where $K_{0}^{{}}=1-\left(  i\hbar^{-1}\mathcal{H}+\Theta\right)  \tau$ and
$K_{1}^{{}}=\sqrt{2\left\langle \Theta\right\rangle \tau}$\ are Kraus
operators, which satisfy the norm conservation condition $\left\langle
K_{0}^{\dag}K_{0}^{{}}+K_{1}^{\dag}K_{1}^{{}}\right\rangle =1+O\left(
\tau^{2}\right)  $ \cite{Daraban_048}, $\hbar$ is the reduced Planck's
constant, $\mathcal{H}^{{}}=\mathcal{H}^{\dag}$ is the system's Hamiltonian,
the positive semi--definite operator $\Theta$ is allowed to depend on $\rho$,
and $\left\langle \Theta\right\rangle =\operatorname{Tr}\left(  \Theta
\rho\right)  $. The corresponding time evolution of the system's state vector
$\left\vert \psi\right\rangle $ can be described using a stochastic
Langevin--Schr\"{o}dinger equation \cite{Grimaudo_033835,Kowalski_167955} [see
Eq. (6) of Ref. \cite{Buks_2604_10562}]. Alternatively, dynamics of the
density operator $\rho$ can be described using a positivity--preserving
\cite{Gorini_821} master equation [see Eq. (5) of Ref. \cite{Buks_2604_10562}].

For the case $\mathcal{H}=0$, and for a fixed operator $\Theta$, Eq.
(\ref{rho(t+tau)}) yields an equation of motion for $\left\langle
\Theta\right\rangle $ given by $\mathrm{d}\left\langle \Theta\right\rangle
/\mathrm{d}t=-2\left\langle \left(  \Theta-\left\langle \Theta\right\rangle
\right)  ^{2}\right\rangle $, and thus for this case the expectation value
$\left\langle \Theta\right\rangle $ monotonically decreases with time. This
observation suggests that the nonlinear term in Eq. (\ref{rho(t+tau)}) can be
used to suppress a given physical property, provided that $\left\langle
\Theta\right\rangle $ quantifies that property.

Two alternative methods to construct the operator $\Theta$ are employed in the
current study. For the first one \cite{Buks_2400036}, which is briefly
reviewed below in this section, the operator $\Theta$ is assumed to be given
by $\Theta=\gamma_{\mathrm{H}}\mathcal{Q}_{\mathrm{H}}+\gamma_{\mathrm{D}%
}\mathcal{Q}_{\mathrm{D}}$, where both rates $\gamma_{\mathrm{H}}$ and
$\gamma_{\mathrm{D}}$ are nonnegative real numbers, and both operators
$\mathcal{Q}_{\mathrm{H}}$ and $\mathcal{Q}_{\mathrm{D}}$\ are Hermitian. The
first term $\gamma_{\mathrm{H}}\mathcal{Q}_{\mathrm{H}}$ gives rise to
thermalization \cite{Grabert_161,Ottinger_052119}, whereas disentanglement is
generated by the second term $\gamma_{\mathrm{D}}\mathcal{Q}_{\mathrm{D}}$.
The second method is discussed in appendix \ref{AppDisCaus}.

The construction of the operator $\mathcal{Q}_{\mathrm{H}}$, which is used for
both methods to generate thermalization, is based on Jaynes' maximum entropy
principle \cite{Jaynes_579,Presse_1115}. The operator $\mathcal{Q}%
_{\mathrm{H}}$ is taken to be given by $\mathcal{Q}_{\mathrm{H}}%
=\beta\mathcal{U}_{\mathrm{H}}$, where $\mathcal{U}_{\mathrm{H}}%
=\mathcal{H}+\beta^{-1}\log\rho$ is the Helmholtz free energy operator,
$\beta=1/\left(  k_{\mathrm{B}}T\right)  $ is the thermal energy inverse,
$k_{\mathrm{B}}$ is the Boltzmann's constant, and $T$ is the temperature. For
the case where $\mathcal{H}^{{}}$ is time independent and $\gamma_{\mathrm{D}%
}=0$ (i.e. no disentanglement), the thermal equilibrium density operator
$\rho_{0}$, which is given by $\rho_{0}=e^{-\beta\mathcal{H}}%
/\operatorname{Tr}\left(  e^{-\beta\mathcal{H}}\right)  $, is a steady state
solution, for which the Helmholtz free energy $\left\langle \mathcal{U}%
_{\mathrm{H}}\right\rangle $ is minimized
\cite{Jaynes_579,Grabert_161,Ottinger_052119,Buks_052217}.

The construction of the disentanglement operator $\mathcal{Q}_{\mathrm{D}}$ is
explained in Ref. \cite{Buks_2400036}. For a multipartite system that is
composed of $N$ subsystems it is given by%
\begin{equation}
\mathcal{Q}_{\mathrm{D}}=\sum_{1\leq s^{\prime}<s^{\prime\prime}\leq
N}\emph{C}_{s^{\prime},s^{\prime\prime}}\;,\label{Q^D}%
\end{equation}
where the operator $\emph{C}_{s^{\prime},s^{\prime\prime}}$\ is defined by%
\begin{equation}
\emph{C}_{s^{\prime},s^{\prime\prime}}=\frac{1}{4}\sum_{n^{\prime}%
=1}^{N_{s^{\prime}}^{2}-1}\sum_{n^{\prime\prime}=1}^{N_{s^{\prime\prime}}%
^{2}-1}\frac{\Lambda_{n^{\prime},n^{\prime\prime}}^{\left(  s^{\prime
},s^{\prime\prime}\right)  }\left\langle \Lambda_{n^{\prime},n^{\prime\prime}%
}^{\left(  s^{\prime},s^{\prime\prime}\right)  }\right\rangle }{1-\left(
\min\left(  N_{s^{\prime}},N_{s^{\prime\prime}}\right)  \right)  ^{-2}%
}\;,\label{C_s_n0,s_n''}%
\end{equation}
and where%
\begin{equation}
\Lambda_{n^{\prime},n^{\prime\prime}}^{\left(  s^{\prime},s^{\prime\prime
}\right)  }=\lambda_{n^{\prime}}^{\left(  s^{\prime}\right)  }\otimes
\lambda_{n^{\prime\prime}}^{\left(  s^{\prime\prime}\right)  }-\left\langle
\lambda_{n^{\prime}}^{\left(  s^{\prime}\right)  }\right\rangle \left\langle
\lambda_{n^{\prime\prime}}^{\left(  s^{\prime\prime}\right)  }\right\rangle
\;.
\end{equation}
The generalized Gell-Mann matrices of subsystem $s$, where $s\in\left\{
1,2,\cdots,N\right\}  $, are denoted by $\lambda_{n}^{\left(  s\right)  }$,
where $n\in\left\{  1,2,\cdots,N_{s}^{2}-1\right\}  $ and $N_{s}$ is the
Hilbert space dimensionality of subsystem $s$ [see Eq. (8.397) of Ref.
\cite{Buks_QMLN}]. The expectation value $\left\langle \emph{C}_{s^{\prime
},s^{\prime\prime}}\right\rangle $, which is bounded by $\left\langle
\emph{C}_{s^{\prime},s^{\prime\prime}}\right\rangle \in\left[  0,1\right]  $,
quantifies the level of correlation between subsystems $s^{\prime}$ and
$s^{\prime\prime}$. Note that $\left\langle \emph{C}_{s^{\prime}%
,s^{\prime\prime}}\right\rangle $ is proportional to the \textit{{linear}}
relative entropy of entanglement [see Eq. (8.440) of Ref. \cite{Buks_QMLN}].
However, the linear relative entropy of entanglement can be used to
approximate the relative entropy of entanglement only in a limited range
\cite{Pauletti_129824}.

Time evolution that is generated by Eq. (\ref{rho(t+tau)}) gives rise to the
suppression of the expectation value $\left\langle \Theta\right\rangle
=\gamma_{\mathrm{H}}\left\langle \mathcal{Q}_{\mathrm{H}}\right\rangle
+\gamma_{\mathrm{D}}\left\langle \mathcal{Q}_{\mathrm{D}}\right\rangle $,
which can be expressed as $\left\langle \Theta\right\rangle =\gamma
_{\mathrm{H}}\beta\left\langle \mathcal{U}_{\mathrm{H},\mathrm{eff}%
}\right\rangle $, where $\mathcal{U}_{\mathrm{H},\mathrm{eff}}=\mathcal{U}%
_{\mathrm{H}}+\beta^{-1}\left(  \gamma_{\mathrm{D}}/\gamma_{\mathrm{H}%
}\right)  \mathcal{Q}_{\mathrm{D}}$ is the effective Helmholtz free energy
operator. The ratio $\gamma_{\mathrm{D}}/\gamma_{\mathrm{H}}$ determines the
relative impact of disentanglement in comparison with thermalization.

\textbf{Macroscopic limit} -- To explore the transition from the
microscopic to the macroscopic realms, we study here the dependency of
disentanglement's impact on system's size. This is done using some many--body
models that describe physical systems composed of $N$ subsystems (particles or
sites) \cite{tasaki2020physics}. For a time independent Hamiltonian
$\mathcal{H}$, The dimensionless parameter $\zeta$ is defined by $\zeta
=\beta\operatorname{spr}\left(  \mathcal{H}\right)  /q_{\mathrm{gs}}$, where
$\operatorname{spr}\left(  \mathcal{H}\right)  $ is the spread of
$\mathcal{H}$ (i.e. the gap between largest and smallest energy eigenvalues),
and $q_{\mathrm{gs}}$ is the ground state value of $\left\langle
\mathcal{Q}_{\mathrm{D}}\right\rangle $. The impact of disentanglement in the
macroscopic limit can be estimated by evaluating the dependency of $\zeta$ on
system's size.

Commonly, many--body models become intractable in the macroscopic limit, i.e.
for $N\gg1$. However exact solutions have been found for some specific models.
Here we focus on three such integrable many--body models: the Lieb-Mattis
antiferromagnet \cite{Lieb_749}, the Affleck-Kennedy-Lieb-Tasa (AKLT) ring
\cite{Affleck_799,Affleck_477}, and the Kitaev chain \cite{Kitaev_131}. For
both the Lieb-Mattis antiferromagnet and the AKLT ring, the system under study
is composed of $N$ distinguishable spin $S$ particles. For the Lieb-Mattis
model $S\in\left\{  1/2,1,3/2,\cdots\right\}  $, whereas $S=1$ for the AKLT
model. Haldane has shown that the low energy properties of a one--dimensional
spin system with an antiferromagnetic interaction depend on the parity of $2S$
\cite{Haldane_464,Haldane_1153}. For odd $2S$ (i.e. for $S\in\left\{
1/2,3/2,\cdots\right\}  $), the ground state commonly becomes degenerate in
the limit $N\rightarrow\infty$. On the other hand, for even $2S$ (i.e. for
$S\in\left\{  1,2,\cdots\right\}  $) the energy gap between the ground state
and the first excited state remains finite in the macroscopic limit of
$N\rightarrow\infty$ (this gap is commonly called the Haldane energy gap)
\cite{Kundu_195146,Zeng_115}. Moreover, correlation functions exponentially
decay for even $2S$. These properties are demonstrated by the exactly solvable
AKLT model \cite{Affleck_799,Affleck_477}. The Kitaev model \cite{Kitaev_131}
allows exploring the macroscopic limit for a system made of indistinguishable
(and spinless) particles. For all three under--study models, the known
solutions allow analytically exploring the impact of disentanglement as a
function of system's size.

\textbf{Lieb-Mattis model} -- Consider a system composed of $2N$
distinguishable spin $1/2$ particles, where $N$ is a positive integer. The
angular momentum vector operator of the $l$'th spin is denoted by
$\mathbf{S}_{l}$. The system is composed of two subsystems labelled as A and
B, and containing $N$ spins each. The Lieb--Mattis Hamiltonian $\mathcal{H}$
is given by \cite{Lieb_749}%
\begin{equation}
\mathcal{H}=\frac{J}{N\hbar}\mathbf{S}_{\mathrm{A}}\cdot\mathbf{S}%
_{\mathrm{B}}\;,
\end{equation}
where $J$ is a real constant, $\mathbf{S}_{\mathrm{A}}=\sum_{l=1}%
^{N}\mathbf{S}_{l}$ and $\mathbf{S}_{\mathrm{B}}=\sum_{l=N+1}^{2N}%
\mathbf{S}_{l}$. To model an antiferromagnet (in the long--range limit), it is
assumed that $J>0$.

For the classical (i.e. fully disentangled) N\'{e}el state $\left\vert
N\left(  \mathbf{\hat{n}}\right)  \right\rangle $, all spins belonging to the
A (B) subsystem are pointing in the $\mathbf{\hat{n}}$ ($-\mathbf{\hat{n}}$)
direction, where $\mathbf{\hat{n}}$ is a unit vector. The energy expectation
value $E_{\mathrm{N}}$ of the N\'{e}el state is given by $E_{\mathrm{N}%
}=\left\langle N\left(  \mathbf{\hat{n}}\right)  \right\vert \mathcal{H}%
\left\vert N\left(  \mathbf{\hat{n}}\right)  \right\rangle =-N\hbar J/4$. The
ground state $\left\vert \psi_{\mathrm{g}}\right\rangle $, which is entangled,
has a lower energy $E_{\mathrm{g}}$ given by $E_{\mathrm{g}}=\left\langle
\psi_{\mathrm{g}}\right\vert \mathcal{H}\left\vert \psi_{\mathrm{g}%
}\right\rangle =-\left(  N\hbar J/4\right)  \left(  1+2/N\right)  $
\cite{Rademaker_032018}, and the state $\left\vert \psi_{\mathrm{g}%
}\right\rangle $\ can be expressed as \cite{Vidal_P01015,Rademaker_013304}%
\begin{equation}
\left\vert \psi_{\mathrm{g}}\right\rangle =\sum_{m=-\frac{N}{2}}^{\frac{N}{2}%
}\frac{\left(  -1\right)  ^{\frac{N}{2}-m}}{\sqrt{N+1}}\left\vert
m,-m\right\rangle \;,
\end{equation}
where $\left\vert m,-m\right\rangle $ is a normalized common eigenvector of
the operators $\mathbf{S}_{\mathrm{A}}^{2}$, $\mathbf{S}_{\mathrm{B}}^{2}$,
$S_{\mathrm{A}z}$ and $S_{\mathrm{B}z}$, with eigenvalues $\left(  N/2\right)
\left(  N/2+1\right)  \hbar^{2}$, $\left(  N/2\right)  \left(  N/2+1\right)
\hbar^{2}$, $m\hbar$ and $-m\hbar$, respectively. Note that the ground state
$\left\vert \psi_{\mathrm{g}}\right\rangle $ is not gaped (i.e. the energy
separation between the ground and first excites states vanishes in the limit
$N\rightarrow\infty$). This property enables a spontaneous symmetry breaking
occurring in the macroscopic limit of $N\rightarrow\infty$ \cite{Beekman_011}.

Consider a subsystem composed of two spins $l^{\prime}$ and $l^{\prime\prime}%
$. The two spins' reduced density operator $\rho_{\left(  l^{\prime}%
,l^{\prime\prime}\right)  }$ is evaluated by partial tracing of the total
density operator $\rho$%
\begin{equation}
\rho_{\left(  l^{\prime},l^{\prime\prime}\right)  }=\operatorname{Tr}%
_{l\notin\left\{  l^{\prime},l^{\prime\prime}\right\}  }\rho\;.
\end{equation}
For the case where $l^{\prime}\in\left\{  1,2,\cdots,N\right\}  $ and
$l^{\prime\prime}\in\left\{  N+1,N+2,\cdots,2N\right\}  $ (i.e. $l^{\prime}$
labels a spin in A, and $l^{\prime\prime}$ labels a spin in B), and for the
ground state $\left\vert \psi_{\mathrm{g}}\right\rangle $, the matrix
representation of $\rho_{\left(  l^{\prime},l^{\prime\prime}\right)  }$\ is
given by [see Eq. (8.468) of \cite{Buks_QMLN}]%
\begin{equation}
\rho_{\left(  l^{\prime},l^{\prime\prime}\right)  }\dot{=}\left(
\begin{array}
[c]{cccc}%
\frac{N-1}{6N} & 0 & 0 & 0\\
0 & \frac{2N+1}{6N} & -\frac{N+2}{6N} & 0\\
0 & -\frac{N+2}{6N} & \frac{2N+1}{6N} & 0\\
0 & 0 & 0 & \frac{N-1}{6N}%
\end{array}
\right)  \;. \label{RDM LM}%
\end{equation}
Note that $\rho_{\left(  l^{\prime},l^{\prime\prime}\right)  }$ is independent
on both $l^{\prime}$ and $l^{\prime\prime}$, $\operatorname{Tr}\rho_{\left(
l^{\prime},l^{\prime\prime}\right)  }=1$ and $\operatorname{Tr}\rho_{\left(
l^{\prime},l^{\prime\prime}\right)  }^{2}=\left(  N^{2}+N+1\right)  /\left(
3N^{2}\right)  $. The partial transpose of $\rho_{\left(  l^{\prime}%
,l^{\prime\prime}\right)  }$ has a negative eigenvalue given by $-1/\left(
2N\right)  $, thus, $\rho_{\left(  l^{\prime},l^{\prime\prime}\right)  }$ is
not separable\ according to the Peres--Horodecki criterion \cite{Peres_1413}.
The two spin ground state entanglement $\tau_{\mathrm{TS}}\equiv\left\langle
\emph{C}_{l^{\prime},l^{\prime\prime}}\right\rangle $ [see Eq.
(\ref{C_s_n0,s_n''})] is given by%
\begin{equation}
\tau_{\mathrm{TS}}=\frac{1}{9}\left(  1+\frac{2}{N}\right)  ^{2}\;.
\label{tau_TS LM}%
\end{equation}
Note that generally (i.e. for an arbitrary two spin 1/2 system),
$\operatorname{Tr}\rho_{\left(  l^{\prime},l^{\prime\prime}\right)  }^{2}$ is
bounded by $\operatorname{Tr}\rho_{\left(  l^{\prime},l^{\prime\prime}\right)
}^{2}\in\left[  1/4,1\right]  $, and $\tau_{\mathrm{TS}}$ is bounded by
$\tau_{\mathrm{TS}}\in\left[  0,1\right]  $. For the ground state $\left\vert
\psi_{\mathrm{g}}\right\rangle $, in the macroscopic limit, i.e. for
$N\rightarrow\infty$, $\operatorname{Tr}\rho_{\left(  l^{\prime}%
,l^{\prime\prime}\right)  }^{2}\rightarrow1/3$ and $\tau_{\mathrm{TS}%
}\rightarrow1/9$.

\begin{figure}[ptb]
\begin{center}
\includegraphics[width=3.2in,keepaspectratio]{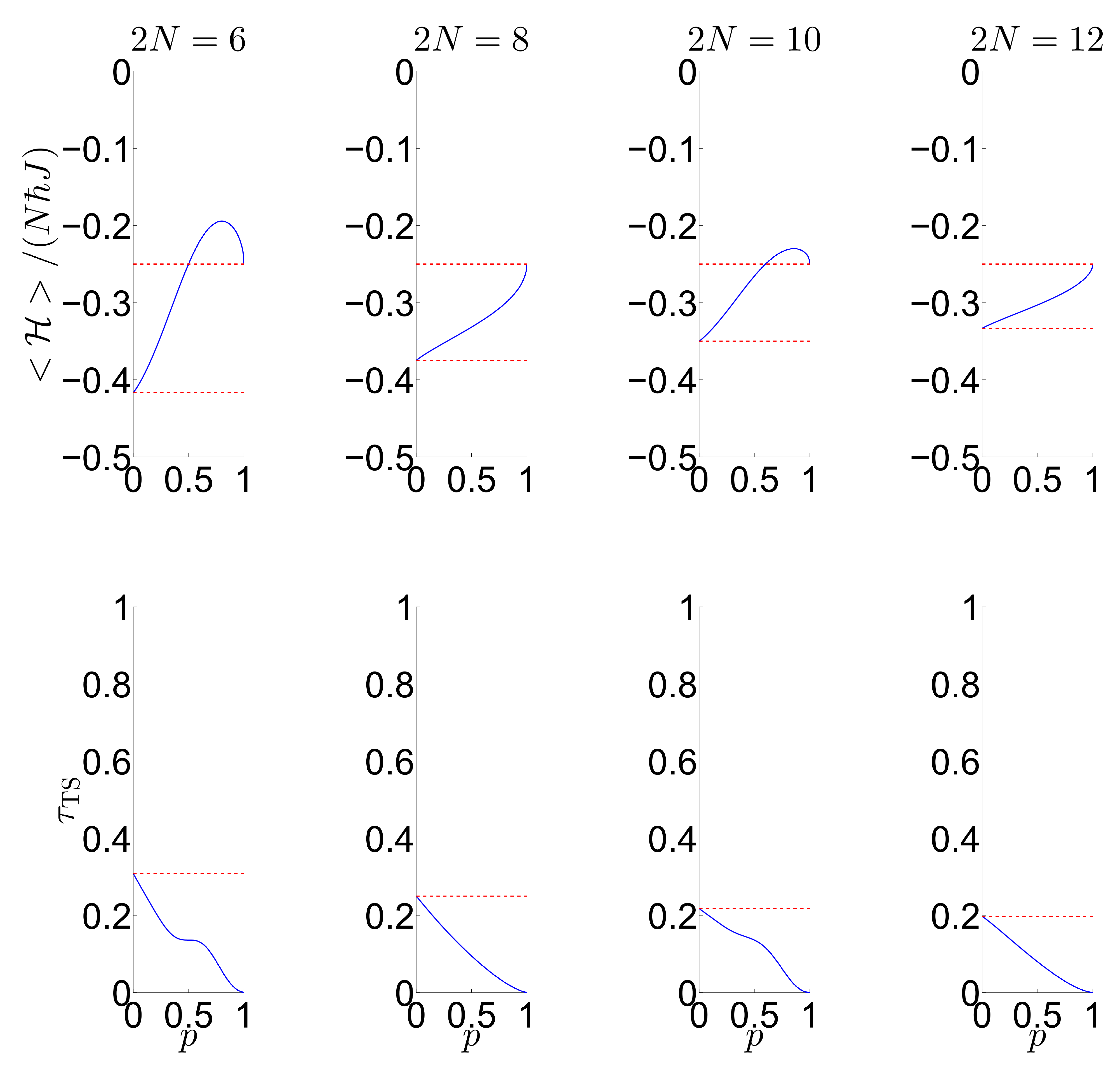}
\end{center}
\caption{{}The Lieb-Mattis model. The energy expectation value $\left\langle
\mathcal{H}\right\rangle $ and the level of entanglement $\tau_{\mathrm{TS}} $
are plotted as a function of $p$ [see Eq. (\ref{psi(p) LM})] for $2N=6$, $8$,
$10$ and $12$.}%
\label{FigLMM}%
\end{figure}

Entanglement \cite{Kaubruegger_2506_10151} is evaluated below for the state
$\left\vert \psi\left(  p\right)  \right\rangle $, which is given by%
\begin{equation}
\left\vert \psi\left(  p\right)  \right\rangle =\frac{\sqrt{1-p}\left\vert
\psi_{\mathrm{g}}\right\rangle +\sqrt{p}\left\vert N\left(  \mathbf{\hat{z}%
}\right)  \right\rangle }{\left\Vert \sqrt{1-p}\left\vert \psi_{\mathrm{g}%
}\right\rangle +\sqrt{p}\left\vert N\left(  \mathbf{\hat{z}}\right)
\right\rangle \right\Vert }\;, \label{psi(p) LM}%
\end{equation}
where $p\in\left[  0,1\right]  $. The state $\left\vert \psi\left(  p\right)
\right\rangle $ represents a weighted average between the ground state
$\left\vert \psi_{\mathrm{g}}\right\rangle $ and the N\'{e}el state
$\left\vert N\left(  \mathbf{\hat{z}}\right)  \right\rangle $ [note that
$\left\vert \psi\left(  p=0\right)  \right\rangle =\left\vert \psi
_{\mathrm{g}}\right\rangle $ and $\left\vert \psi\left(  p=1\right)
\right\rangle =\left\vert N\left(  \mathbf{\hat{z}}\right)  \right\rangle $)].
The plots in the top (first) row of Fig. \ref{FigLMM} display the normalized
energy expectation value $\left\langle \psi\left(  p\right)  \right\vert
\mathcal{H}\left\vert \psi\left(  p\right)  \right\rangle /\left(  N\hbar
J\right)  $ as a function of $p$. The normalized energy expectation values of
$-1/4$,\ for the N\'{e}el state, and of $-\left(  1/4\right)  \left(
1+2/N\right)  $, for the ground state, are represented by overlaid red dashed
horizontal lines. The plots in the second row of Fig. \ref{FigLMM} display the
entanglement $\tau_{\mathrm{TS}}$ between two spins $l^{\prime}$ and
$l^{\prime\prime}$ (where $l^{\prime}\in\left\{  1,2,\cdots,N\right\}  $ and
$l^{\prime\prime}\in\left\{  N+1,N+2,\cdots,2N\right\}  $) as a function of
$p$. The overlaid red dashed horizontal lines in the second row of Fig.
\ref{FigLMM} represent the value given by Eq. (\ref{tau_TS LM}) for the ground
state entanglement $\tau_{\mathrm{TS}}$. The number $N$ for the first (left),
second, third and forth column in Fig. \ref{FigLMM} is $3$, $4$, $5$ and $6$, respectively.

As was discussed above, the nonlinear term in the equations of motion
suppresses the expectation value $\left\langle \Theta\right\rangle
=\gamma_{\mathrm{H}}\left\langle \mathcal{Q}_{\mathrm{H}}\right\rangle
+\gamma_{\mathrm{D}}\left\langle \mathcal{Q}_{\mathrm{D}}\right\rangle $. For
the Lieb-Mattis model at low temperatures it is expected that the disentanglement term
$\gamma_{\mathrm{D}}\left\langle \mathcal{Q}_{\mathrm{D}}\right\rangle $
becomes dominant in the macroscopic limit, because in this limit its ground
state value is proportional to $N^{2}$ [see Eqs. (\ref{Q^D}) and
(\ref{tau_TS LM})].

\textbf{AKLT model} -- The impact of disentanglement on a gaped ground state
can be explored using the AKLT model. Consider a one-dimensional array
composed of $L$ spin $S=1$ localized particles. The angular momentum vector
operator of the $l$'th spin is denoted by $\mathbf{S}_{l}=\left(
S_{l,x},S_{l,y},S_{l,z}\right)  $, where $l\in\left\{  1,2,\cdots,L\right\}
$. The Hilbert space of the single spin occupying site $l$, where
$l\in\left\{  1,2,\cdots,L\right\}  $, is spanned by an orthonormal basis
given by $\left\{  \left\vert -;l\right\rangle ,\left\vert 0;l\right\rangle
,\left\vert 1;l\right\rangle \right\}  $. The $q$ -- deformed matrix product
state \cite{Maekawa_031901} is defined by $\left\vert \psi_{q}\right\rangle
=\operatorname{Tr}\left(  \mu_{1}\mu_{2}\cdots\mu_{L}\right)  $,\ where the
matrices $\mu_{l}$ are given by \cite{Klumper_281}%
\begin{equation}
\mu_{l}=\left(
\begin{array}
[c]{cc}%
q^{-1}\left\vert 0;l\right\rangle  & -\sqrt{q^{{}}+q^{-1}}\left\vert
+;l\right\rangle \\
\sqrt{q^{{}}+q^{-1}}\left\vert -;l\right\rangle  & -q\left\vert
0;l\right\rangle
\end{array}
\right)  \;, \label{mu_l qD MPS}%
\end{equation}
and $q$ is real (note that $\left\vert \psi_{q}\right\rangle $ is not
normalized). The state $\left\vert \psi_{q}\right\rangle $ for the case $q=1$
is the ground state of the AKLT Hamiltonian $\mathcal{H}_{\mathrm{AKLT}}$,
which is given by%
\begin{equation}
\mathcal{H}_{\mathrm{AKLT}}=\sum_{l=1}^{L}P_{2}^{\left(  l,l+1\right)  }\;,
\label{H AKLT}%
\end{equation}
where $P_{2}^{\left(  l,l+1\right)  }$\ is the projection operator
corresponding to the pair of $l$ and $l+1$ spins, and to the quantum number
$S=2$ [see Eq. (18.569) of \cite{Buks_QMLN}]. For the last term $l=L$, the
projection operator $P_{2}^{\left(  L,L+1\right)  }$ represents the operator
$P_{2}^{\left(  L,1\right)  }$. In the macroscopic limit, i.e. for $L\gg1$,
the Haldane energy gap \cite{Haldane_464,Haldane_1153} $\Delta_{\mathrm{AKLT}%
}$ for the AKLT Hamiltonian (\ref{H AKLT}) is $\Delta_{\mathrm{AKLT}}%
\simeq0.35$ \cite{Garcia_245118,auerbach2012interacting}. While the AKLT
ground state $\left\vert \psi_{q=1}\right\rangle $ is entangled, for
$q=0$\ the state $\left\vert \psi_{q}\right\rangle $ becomes fully disentangled.

\begin{figure}[ptb]
\begin{center}
\includegraphics[width=3.2in,keepaspectratio]{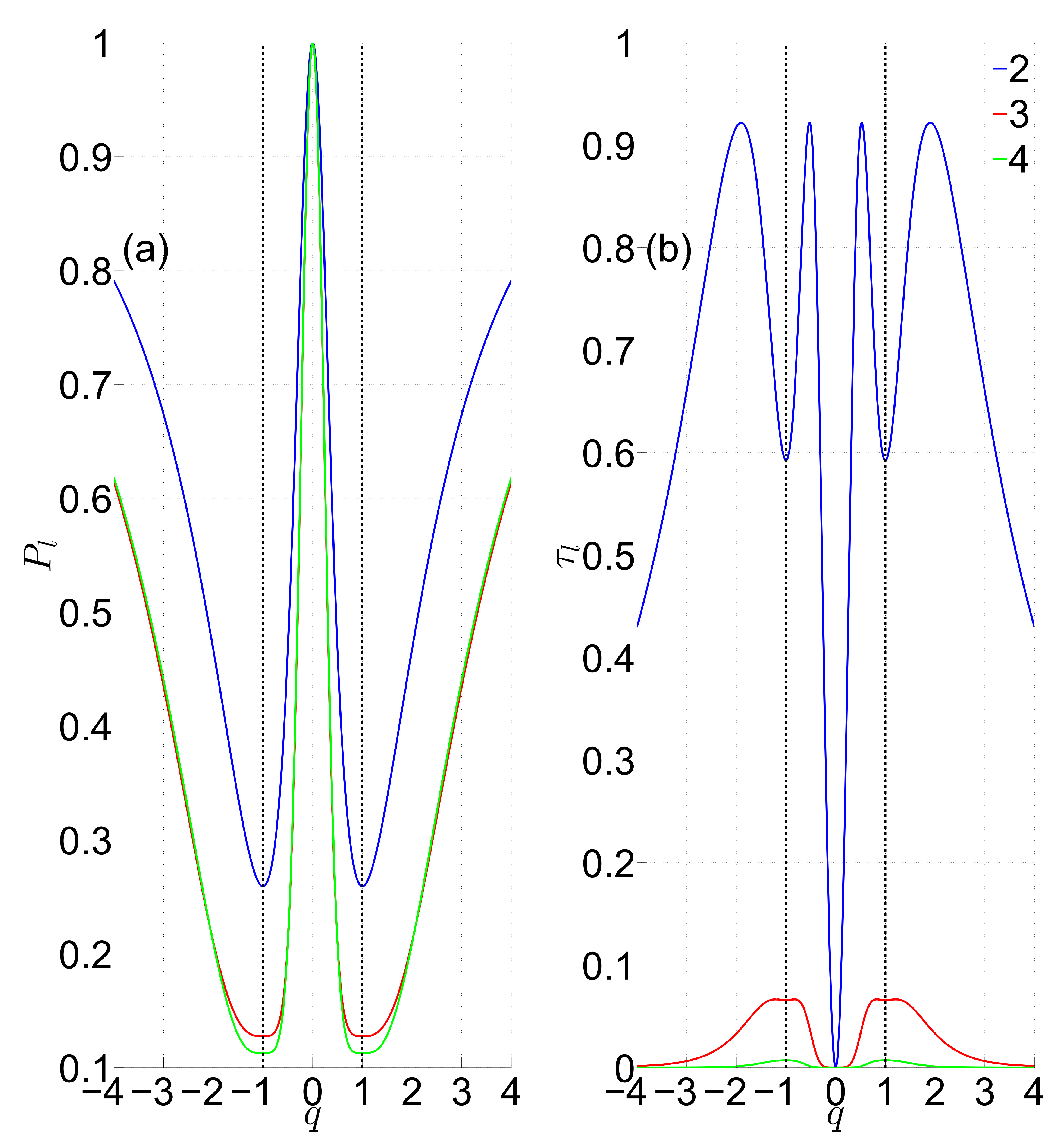}
\end{center}
\caption{{}AKLT model. Both (a) purity $P_{l}=\operatorname{Tr}\rho_{\left(
1,l\right)  }^{2}$ and (b) entanglement $\tau_{l}$ are plotted as a function
of $q$. The colors blue, red and green are used for the site spacing $l$
values of 2, 3 and 4, respectively. The overlaid black dashed vertical lines
label the values $q=\pm1$ (AKLT ground state).}%
\label{FigAKLT}%
\end{figure}

The reduced density matrix corresponding to the subsystem composed of the
first and the $l$'th sites is denoted by $\rho_{\left(  1,l\right)  }$. The
plot in Fig. \ref{FigAKLT}(a) shows the subsystem purity $\operatorname{Tr}%
\rho_{\left(  1,l\right)  }^{2}$ as a function of $q$. Note that
$\operatorname{Tr}\rho_{\left(  1,l\right)  }^{2}=\operatorname{Tr}\left(
\mathcal{M}_{\mathrm{W}}^{\dag}\mathcal{M}_{\mathrm{W}}^{{}}\right)  $, where
$\mathcal{M}_{\mathrm{W}}$ is the subsystem Weyl matrix [see Eq. (8.339) of
Ref. \cite{Buks_QMLN}]. The entanglement \cite{Zhang_062,Santos_37005} between
the first and the $l$'th sites, which is denoted by $\tau_{l}\equiv
\left\langle \emph{C}_{1.l}\right\rangle $ [see Eq. (\ref{C_s_n0,s_n''})], is
plotted in Fig. \ref{FigAKLT}(b) as a function of $q$. For the case $q=1$,
i.e. for the AKLT ground state, and in the limit $L\gg1$, the following holds
$\operatorname{Tr}\rho_{\left(  1,l\right)  }^{2}=1/9+\left(  4/27\right)
\left(  1/9\right)  ^{l-2}$ and [see Eq. (18.1062)) of Ref. \cite{Buks_QMLN},
and note that $l\geq2$]%
\begin{equation}
\tau_{l}=\frac{16}{27}\left(  \frac{1}{9}\right)  ^{l-2}\;. \label{tau_l AKLT}%
\end{equation}
Thus both $\operatorname{Tr}\rho_{\left(  1,l\right)  }^{2}$ and $\tau_{l}%
$\ exponentially decrease as a function of site spacing $l$ to their lowest
allowed values ($1/9$ and $0$, respectively). This observation suggests that
entanglement in the AKLT ground state is local, and remote spins are uncorrelated.

\begin{figure}[ptb]
\begin{center}
\includegraphics[width=3.2in,keepaspectratio]{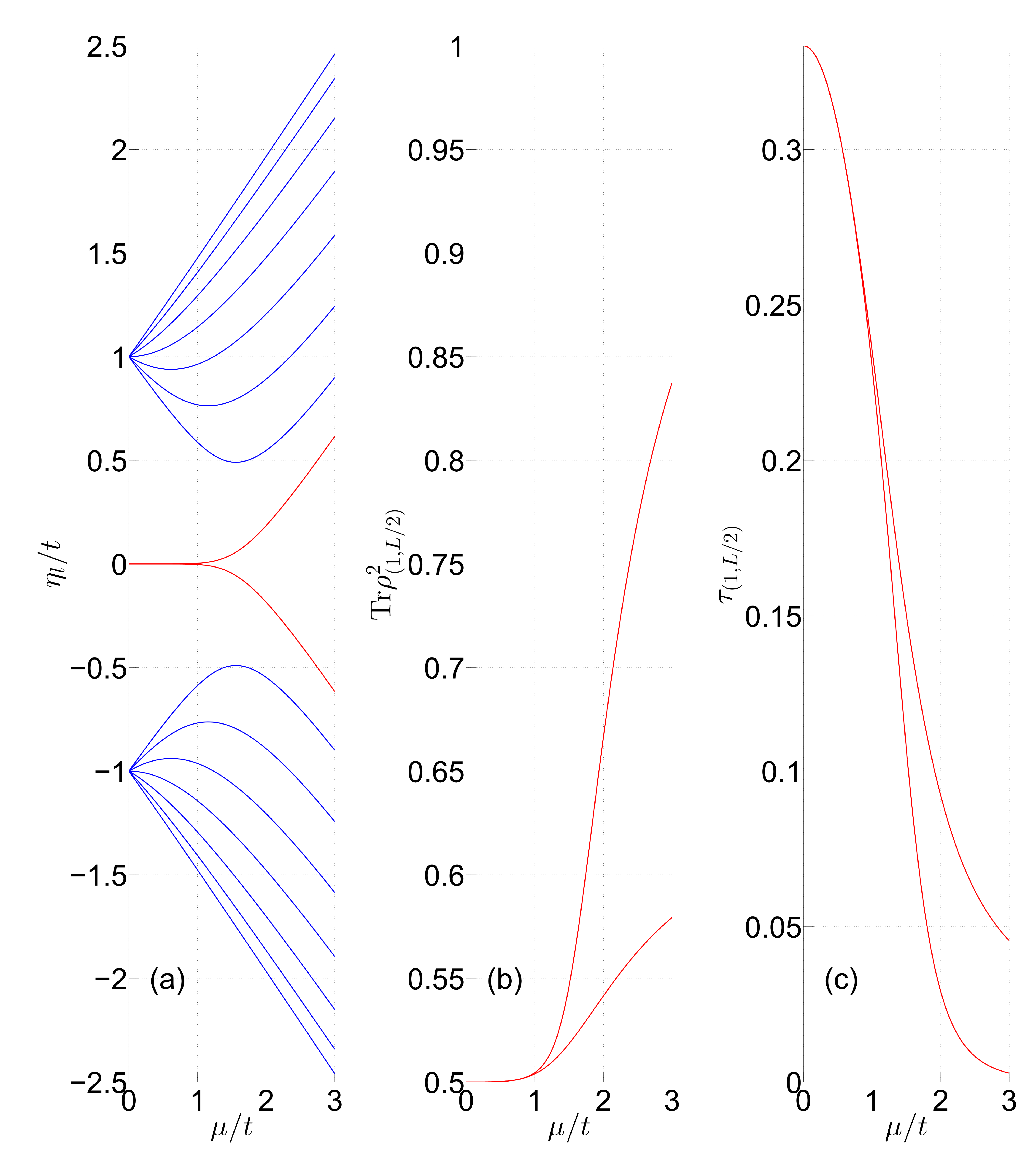}
\end{center}
\caption{{}Kitaev chain. The (a) Bogoliubov eigenvalues $\eta_{l}/t$, (b) two
sites purity\ $\operatorname{Tr}\rho_{\left(  1,L/2\right)  }^{2}$ and (c) the
two sites level of entanglement $\tau_{\left(  1,L/2\right)  }$ are plotted as
as a function of the ratio $\mu/t$, for the case $L=8$.}%
\label{FigKC}%
\end{figure}

\textbf{Kitaev model} -- The effect of disentanglement on indistinguishable
particles can be studied using the Kitaev model. Consider a one--dimensional
array containing $L$ sites that are occupied by \textit{spinless} Fermions.
The creation and annihilation operators corresponding to site $l\in\left\{
1,2,\cdots,L\right\}  $ are denoted by $a_{l}^{\dag}$ and $a_{l}^{{}}$,
respectively. The operators $a_{l}^{\dag}$ and $a_{l}^{{}}$ satisfy Fermionic
anti-commutation relations. The Kitaev chain Hamiltonian $\mathcal{H}%
_{\mathrm{K}}$ is given by%
\begin{align}
\mathcal{H}_{\mathrm{K}}  &  =-t\sum_{l=1}^{L-1}\left(  a_{l}^{\dag}%
a_{l+1}^{{}}+a_{l+1}^{\dag}a_{l}^{{}}-a_{l}^{\dag}a_{l+1}^{\dag}-a_{l+1}^{{}%
}a_{l}^{{}}\right) \nonumber\\
&  -\mu\sum_{l=1}^{L}\left(  a_{l}^{\dag}a_{l}^{{}}-\frac{1}{2}\right)
\ ,\nonumber\\
&  \label{H K}%
\end{align}
where both $t$ and $\mu$ are real constants. The Hamiltonian $\mathcal{H}%
_{\mathrm{K}}$ can be diagonalized using a Bogoliubov transformation. The plot
in Fig.\ref{FigKC}(a) displays the normalized Bogoliubov eigenvalues $\eta
_{l}/t$ , where $l\in\left(  1,2,\cdots,2L\right)  $, as a function of the
ratio $\mu/t$. The two nearly vanishing eigenvalues in the region
$\mu/t\lesssim1$ , which are red colored in Fig. \ref{FigKC}(a), are commonly
referred to as Majorana zero modes \cite{Sajith_184509}. For the case $\mu=0$,
the ground state, which is doubly degenerate, has energy given by $-t\left(
L-1\right)  $.

The two sites $l^{\prime}$ and $l^{\prime\prime}$ reduced density operator is
denoted by $\rho_{\left(  l^{\prime},l^{\prime\prime}\right)  }$. For the case
$\mu=0$ the corresponding purity, which is given by $\operatorname{Tr}%
\rho_{\left(  l^{\prime},l^{\prime\prime}\right)  }^{2}=1/2$ [see Eq. (18.557)
of Ref. \cite{Buks_QMLN}], is independent on both site indices $l^{\prime}$
and $l^{\prime\prime}$. Similarly, the level of entanglement between sites
$l^{\prime}$ and $l^{\prime\prime}$
\cite{Fromholz_085136,Chitov_085131,Granet_133,Kuboki_2510_20312,Borla_148,Medina_214512}%
, which is denoted by $\tau_{\left(  l^{\prime},l^{\prime\prime}\right)
}\equiv\left\langle \emph{C}_{l^{\prime},l^{\prime\prime}}\right\rangle $ [see
Eq. (\ref{C_s_n0,s_n''})] is independent on both $l^{\prime}$ and
$l^{\prime\prime}$ for the case $\mu=0$. The dependency of the purity
$\operatorname{Tr}\rho_{\left(  1,L/2\right)  }^{2}$ and entanglement
$\tau_{\left(  1,L/2\right)  }$ on the ratio $\mu/t$ is shown in Fig.
\ref{FigKC}(b) and (c), respectively, for the case $L=8$. For $\mu=0$ and for
arbitrary $L$%
\begin{equation}
\tau_{\left(  l^{\prime},l^{\prime\prime}\right)  }=\frac{1}{3}\ ,
\label{tau lp,lpp K}%
\end{equation}
and thus for this case the Kitaev ground state is non--locally entangled.

\textbf{Discussion} -- In the macroscopic limit (i.e. $N\gg1$ for the
Lieb-Mattis model, and $L\gg1$ for the AKLT and Kitaev ones), the parameter
$q_{\mathrm{gs}}$ (the ground state value of $\left\langle \mathcal{Q}%
_{\mathrm{D}}\right\rangle $) scales as $N^{2}$ for the case of the
Lieb-Mattis model [see Eq. (\ref{tau_TS LM})], as $L$ for the AKLT model [see
Eq. (\ref{tau_l AKLT})], and as $L^{2}$ for the Kitaev model with $\mu=0$ [see
Eq. (\ref{tau lp,lpp K})]. Thus, both the Lieb-Mattis and Kitaev (for $\mu=0$)
ground states become unstable in the macroscopic limit (the dimensionless
parameter $\zeta\rightarrow0$ in this limit for the Kitaev model). In
contrast, the possibility that the AKLT ground state remains stable in the
macroscopic limit is not excluded.

The expectation value of $\left\langle \mathcal{Q}_{\mathrm{D}}\right\rangle $
associated with the AKLT ground state obeys an area law
\cite{Eisert_277,vanAcoleyen_170501}, and, consequently, correlation functions
exponentially decay as a function of spacing between sites for this state [see
Eq. (\ref{tau_l AKLT})]. In contrast, long range correlations occur for both
Lieb-Mattis and Kitaev (for $\mu=0$) ground states. The examples presented in
the current study, which are based on many--body models having known exact
solutions, suggest a possible connection between the stability of macroscopic
entangled states against the impact of disentanglement, and between the
correlation length associated with the given state. However, further study is
needed to incisively determine under what conditions a general macroscopic
entangled state is kept stable against disentanglement.

\textbf{Summary} -- The proposed nonlinear extension yields time
evolution, which incorporates the processes of thermalization and collapse
with unitary time evolution. In the current study, the
transition from the microscopic limit to the macroscopic one is explored. Our
findings demonstrate that, for some cases, the impact of disentanglement can
be significant in the macroscopic limit, even when that impact is kept
arbitrary small in the microscopic limit. In other words, for these cases, the
spontaneous disentanglement hypothesis provides a possible way to reconcile
between the quantumness of the microscopic realm, and the classicalness of the
macroscopic one. The hypothesis is falsifiable, since it yields predictions
that are distinguishable from what is obtained using alternative models.
However, deriving predictions that are applicable in the macroscopic limit,
and that are experimentally testable, is challenging, since commonly
many--body models become intractable in the macroscopic limit (even without
implementing disentanglement, which further complicates calculations).

\textbf{Acknowledgments} -- Useful discussions with Nathan Argaman are acknowledged.

\appendix

\section{Disentanglement and causality}

\label{AppDisCaus}

The nonlinear equation of motion (\ref{rho(t+tau)}) is constructed
in the main text by assuming that the operator $\Theta$ is given by
$\Theta=\gamma_{\mathrm{H}}\mathcal{Q}_{\mathrm{H}}+\gamma_{\mathrm{D}%
}\mathcal{Q}_{\mathrm{D}}$ (see Ref. \cite{Buks_2400036}). For this
construction method, however, and for a spatially extended quantum system, the
resultant dynamics may give rise to an inconsistency with the Einstein's
causality principle \cite{Gisin_1,Rembielinski_012027}. Under some appropriate
conditions, this inconsistency can be avoided by implementing an alternative
method to construct the nonlinear equation of motion
\cite{Buks_2604_10562}. For this alternative method, stability of entangled quantum
states in the macroscopic limit is discussed in this appendix.

In the first step of the alternative method, it is assumed that $\Theta
=\gamma_{\mathrm{H}}\mathcal{Q}_{\mathrm{H}}$ (recall that $\mathcal{Q}%
_{\mathrm{H}}=\beta\mathcal{U}_{\mathrm{H}}$, where $\mathcal{U}_{\mathrm{H}}$
is the Helmholtz free energy operator). While the operator $\Theta$ is assumed
to have only a thermalization term, disentanglement is introduced in the next
step by enforcing an additional constraint \cite{Buks_2604_10562}.

Consider the case where the quantum system under study is composed of two
subsystems, which are labeled by the letters $\mathrm{a}$ and $\mathrm{b}$,
respectively. The bipartite mutual information $\mathcal{I}_{\mathrm{a}%
,\mathrm{b}}$ is expresses as the relative entropy $\sigma\left(  \rho_{{}%
}\parallel\rho_{\mathrm{a}}\otimes\rho_{\mathrm{b}}\right)  $ between the
density operators $\rho$ and $\rho_{\mathrm{a}}\otimes\rho_{\mathrm{b}}$
\cite{Wehrl_221}%
\begin{equation}
\mathcal{I}_{\mathrm{a},\mathrm{b}}=\sigma_{\mathrm{a}}+\sigma_{\mathrm{b}%
}-\sigma_{{}}=\sigma\left(  \rho\parallel\rho_{\mathrm{a}}\otimes
\rho_{\mathrm{b}}\right)  \;,
\end{equation}
where the subsystems' entropies $\sigma_{\mathrm{a}}$ and $\sigma_{\mathrm{b}%
}$ are given by $\sigma_{\mathrm{a}}=-\operatorname{Tr}\left(  \rho
_{\mathrm{a}}\log\rho_{\mathrm{a}}\right)  $ and $\sigma_{\mathrm{b}%
}=-\operatorname{Tr}\left(  \rho_{\mathrm{b}}\log\rho_{\mathrm{b}}\right)  $,
respectively, and where the subsystems' reduced density operators
$\rho_{\mathrm{a}}$ and $\rho_{\mathrm{b}}$\ are derived from the density
operator $\rho$ of the composed system by partial tracing, i.e. $\rho
_{\mathrm{a}}=\operatorname{Tr}_{\mathrm{b}}\rho$ and $\rho_{\mathrm{b}%
}=\operatorname{Tr}_{\mathrm{a}}\rho$. It has been shown that relative entropy
and quantum mutual information can be used to formulate entanglement area laws
\cite{Wolf_070502}, and both the second \cite{Sagawa_125} and third
\cite{Floerchinger_052117} laws of thermodynamics. The Klein subadditivity
inequality \cite{Klein_767,Wehrl_221}, which states that $\sigma\left(
\rho^{\prime}\parallel\rho^{\prime\prime}\right)  \equiv\operatorname{Tr}%
\left(  \rho^{\prime}\left(  \log\rho^{\prime}-\log\rho^{\prime\prime}\right)
\right)  \geq0$, implies that $\sigma_{\mathrm{a}}+\sigma_{\mathrm{b}}%
-\sigma_{{}}\geq0$, and equality holds (i.e. $\sigma_{{}}=\sigma_{\mathrm{a}%
}+\sigma_{\mathrm{b}}$) if and only if $\rho_{{}}=\rho_{\mathrm{a}}\otimes
\rho_{\mathrm{b}}$.

For a bipartite system, the total entropy $\sigma$ can be expressed as
$\sigma_{{}}=\sigma_{\mathrm{a}}+\sigma_{\mathrm{b}}-\mathcal{I}%
_{\mathrm{a},\mathrm{b}}$. The subadditivity inequality implies that the
disentangled state $\rho_{\mathrm{a}}\otimes\rho_{\mathrm{b}}$ maximizes the
entropy $\sigma$, under the constraints that both subsystems' reduced density
operators $\rho_{\mathrm{a}}$ and $\rho_{\mathrm{b}}$ are fixed. The mutual
information minimum principle is implemented by enforcing these constraints.
As is explained in Ref. \cite{Buks_2604_10562}, this can be done using the
method of Lagrange multipliers. With these enforced constraints, the process
of thermalization gives rise to the suppression of the quantum mutual
information, which, in turn generates disentanglement between subsystems
$\mathrm{a}$ and $\mathrm{b}$. However, the constraints (that fix both
$\rho_{\mathrm{a}}$ and $\rho_{\mathrm{b}}$) ensure that this process of
disentanglement has no impact on any single subsystem property. Hence, a
conflict with causality, which can occur when the composed quantum system is
spatially extended, is avoided.

Consider the case where the system's Hamiltonian
$\mathcal{H}$ is time-independent, and can be expressed as $\mathcal{H}=\mathcal{H}_{\mathrm{a}}+\mathcal{H}%
_{\mathrm{b}}+V$, where $\mathcal{H}_{\mathrm{a}}$ and $\mathcal{H}%
_{\mathrm{b}}$ are Hamiltonians of subsystems a and b, respectively, and the
term $V$ represents interaction between the two subsystems. For this case, the
entanglement area law \cite{Wolf_070502} yields [recall that $\rho
_{0}=e^{-\beta\mathcal{H}}/\operatorname{Tr}\left(  e^{-\beta\mathcal{H}%
}\right)  $ is the thermal equilibrium density operator]%
\begin{equation}
\sigma\left(  \rho_{0}\parallel\rho_{\mathrm{a}}\otimes\rho_{\mathrm{b}%
}\right)  \leq2\beta\left\Vert V\right\Vert _{\infty}\ ,
\label{entanglement area law}%
\end{equation}
where the Schatten norm $\left\Vert V\right\Vert _{\infty}$ is defined by
$\left\Vert V\right\Vert _{\infty}=\lim_{p\rightarrow\infty}\left(
\operatorname{Tr}\left(  \left\vert O\right\vert ^{p}\right)  \right)  ^{1/p}$
[see Eq. (17.232) of Ref. \cite{Buks_QMLN}].

The inequality (\ref{entanglement area law}) can be used to estimate the
stability of entangled states in the macroscopic limit. For the integrable
models under study, the Schatten norm $\left\Vert V\right\Vert _{\infty}$
scales as $N^{2}$ for the case of the Lieb-Mattis model, as $L$ for the AKLT
model, and as $L^{2}$ for the Kitaev model with $\mu=0$. Thus, for both the
Lieb-Mattis and Kitaev (for $\mu=0$) models, instability in the macroscopic
limit cannot be excluded.

\bibliographystyle{ieeepes}
\bibliography{acompat,Eyal_Bib}

\end{document}